\documentclass[a4paper]{article}
\usepackage{epsfig,amsmath,amsfonts,amssymb,setspace,multirow,textcomp}
\usepackage[T1]{fontenc}
\usepackage{color}
\makeatletter
\renewcommand{\@oddhead}{\textit{Advances in Astronomy and Space Physics} \hfil}
\renewcommand{\@evenfoot}{\hfil \thepage \hfil}
\renewcommand{\@oddfoot}{\hfil \thepage \hfil}
\makeatother

\renewenvironment{thebibliography}[1]{\begin{oldthebibliography}{#1}\setlength{\parskip}{0ex}\setlength{\itemsep}{0ex}}{\end{oldthebibliography}}

\begin{document}
\fontsize{11}{11}\selectfont 
\title{Detection of the rapid variability in the Q2237+0305 quasar}
\author{\textsl{L.\,A.~Berdina$^{1,2}$, V.\,S.~Tsvetkova$^{1,2}$}}
\date{\vspace*{-6ex}}
\maketitle
\begin{center} {\small $^{1}$Institute of Radio Astronomy of Nat.Ac.Sci. of Ukraine, 4 Mystetstv, 61002 Kharkov, Ukraine\\
$^{2}$Institute of Astronomy of Kharkov National University,
Sumskaya 35, 61022 Kharkov, Ukraine\\
{\tt laberdina@gmail.com}}
\end{center}

\begin{abstract}
Rapid intrinsic variability has been detected for the first time in the Einstein Cross QSO 2237+0305, a radio quiet flat spectrum quasar at z=1.7 that is quadruply lensed by a foreground galaxy at z=0.04. The observed short-period event at the time scale of several days and with amplitudes of about 0.1-0.2 mag can be traced in the light
curves of the 2004 observing season for all macroimages and in all the three filters (
$V$ , $R$ and $I$). The accuracy of the existing estimates of the time delays in Q2237+0305 is insufficient to either confirm or disprove the estimations of time delays based on the lens models of this system, especially taking into account the presence of strong microlensing events. The detected short-period variations in the Q2237+0305 light curves have made it possible to obtain new estimates of the time delays, which are more accurate as compared to the earlier determinations by other authors.\\[1ex]
\end{abstract}

\section*{\sc introduction}
\indent \indent Quasars are known to be variable objects, which change their brightness at a wide range of time scales, - from several hours to several years. In gravitational lens systems, we have an opportunity to observe the quasars intrinsic brightness variations repeated in all macroimages and shifted in time depending on the system geometry, distribution of the gravitational lens potential and the adopted cosmological model.  An importance of measuring these time shifts in gravitationally lensed quasars was noted for the first time by Refsdal \cite{refsdal64}, who has shown that such measurements make it possible to determine the Hubble constant without involving the intermediate distance standards. There are also other important applications of the gravitational time delays in astrophysics and cosmology, such as the study of matter distribution at different spatial scales in the Universe, including the dark matter, investigation of the spatial structure of lensing galaxies, and others.

In order to be useful for determination of the Hubble constant, the time delays should be measured with a relative error of the order of 1 \%
\cite{kochanek04}. Until recently, it was a problem to achieve this accuracy, but however, due to the great interest to this problem, significant changes have been occurred during the last years (\cite{courbin11}, \cite{eulaers13}, \cite{tewes13a}, \cite{tewes13b}, \cite{kumar13}, \cite{bonvin16}).

The accuracy of measuring the time delays between images of a gravitationally lensed quasar depends on many factors. Some of them are laid at the initial stage and are related to the initial observational data acquisition: photometry errors, irregular sampling of the light curves, the presence of large seasonal gaps, etc. It seems possible to minimize a number of the above-mentioned factors by improving the data acquisition procedure. However, there are a number of other factors, which cannot be controlled by the observer, such as: small amplitudes of variations in the intrinsic quasar brightness, microlensing events caused by motions of objects populating the lensing galaxy near the light path between a quasar, macroimage and an observer.

One of the most difficult factors among the above-listed ones is variable microlensing, which affects the accuracy of the time delays measurements most of all. In particular, in the case of an extended source, variable microlensing produced by the moving objects in the lensing galaxy will differently magnify different parts of the source in quasar images, thus differently distorting the shapes and amplitudes of the light curves of macroimages \cite{barkana97}. The detailed theoretic analysis of the peculiarities of gravitational focusing of time-variable and extended sources has been made for the first time in the work \cite{minakov03}. It has been shown that for such sources, not only the spatial redistribution of radiation, but also redistribution in time may occur in gravitational focusing of radiation. This leads to the fact that the quasar light curve observed in different macroimages will be distorted in different ways. Significant effects of variable microlensing of an extended source on the accuracy of measuring the time delays are also discussed in the most recent works (\cite{sluse14}, \cite{tie18}).

In this paper, detection of a rapid brightness variations in Q2237+0305 is reported, which is made from analysis of the light curves presented by Dudinov \cite{dudinov11}. Some preliminary results of determining the time delays in Q2237+0305 made on the basis of the detected short-period events are presented.

\section*{\sc short-period flux variations in the q2237+0305 light curves}

\indent \indent The characteristics of flux variations in quasars are under scrutiny of the astronomical community because of their importance for understanding the mechanisms of their variability (see, e.g. \cite{cristiani97},\cite{giveon99}, \cite{vanden04}, \cite{magdis06}, \cite{wilhite08}, \cite{gopal13}). Quasars are known to change their flux in a wide range of time scales (from days to several years), with the typical amplitudes growing towards the larger time lags and shorter wavelengths (\cite{wilhite08}, \cite{webb00}). The variability on the time scales of days and hours is of a particular interest, since it is diagnostic in distinguishing between various mechanisms of the quasar variability and constraining the size of a quasar region responsible for variability.

The rapid (day-to-day or even intra day) variability in some radio-quiet quasars (RQQs) have been reported by many authors (e.g. \cite{gopal13}, \cite{diego98}, \cite{chand15} and others). In particular, the evidences for the short-period variability were presented in the paper \cite{jang97} for two out of 10 radio-quiet QSOs and for 6 out of 7 radio-loud QSOs. The results of monitoring the short-period optical variability in broad absorption lines on timescales of about 1 hr in QSOs 0846+156 and 0856+172 were demonstrated in \cite{anupama98}.

We revealed the short-period flux variations in the radio-quiet quasar Q2237+0305 in the process of examining its light curves reported by \cite{dudinov11}. Observations of Q2237+0305 covering the time period from 2001 to 2008 were carried out with the 1.5-m telescope of the high-altitude Maidanak Observatory (Central Asia, Uzbekistan) with a scientific BroCam CCD camera. The observations were carried out in filters $V$, $R$, and $I$ of the Johnson-Cousins photometric system under very good seeing conditions (FWHM of the reference star is less than 1"). The largest number of images of Q2237 + 0305 has been obtained in filter $R$, with the sampling rate of 1 to 3 days.

\begin{figure}[!h]
\centering
\begin{minipage}[t]{.47\linewidth}
\centering
\epsfig{file = 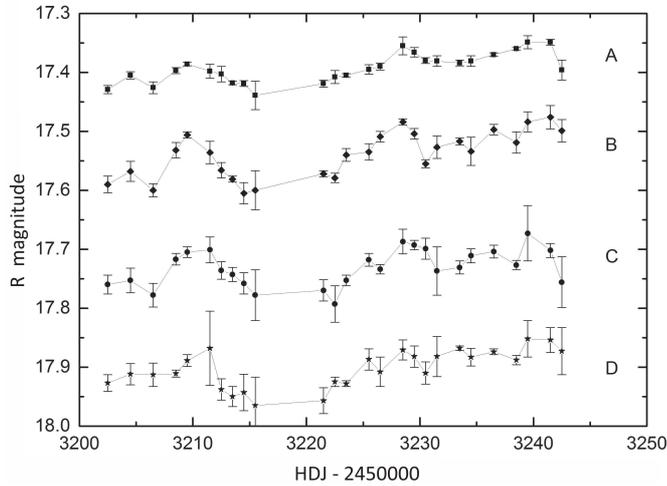,width = 1.03\linewidth}
\caption{The light curves of Q2237+0305 (season 2004) in filter $R$. The magnitudes (along the vertical axis) are shifted arbitrarily for better view; Julian dates - along the horizontal axis.}\label{fig1}
\end{minipage}
\hfill
\end{figure}

The short-period event in the Q2237+0305 system was found in the light curves of the 2004 season. The event is seen in all the four image components of the quasar but the $D$ component, where its detection is rather uncertain because of a lower signal-to-noise ratio. This is clearly seen in Fig.~\ref{fig1} where the results of observation in filter $R$ are presented. The maximum of this event is concentrated near the Julian date 2453210. The data are available at http://www.astron.kharkov.ua/databases/index.html.

\begin{figure}[!h]
\centering
\begin{minipage}[t]{.47\linewidth}
\centering
\epsfig{file = 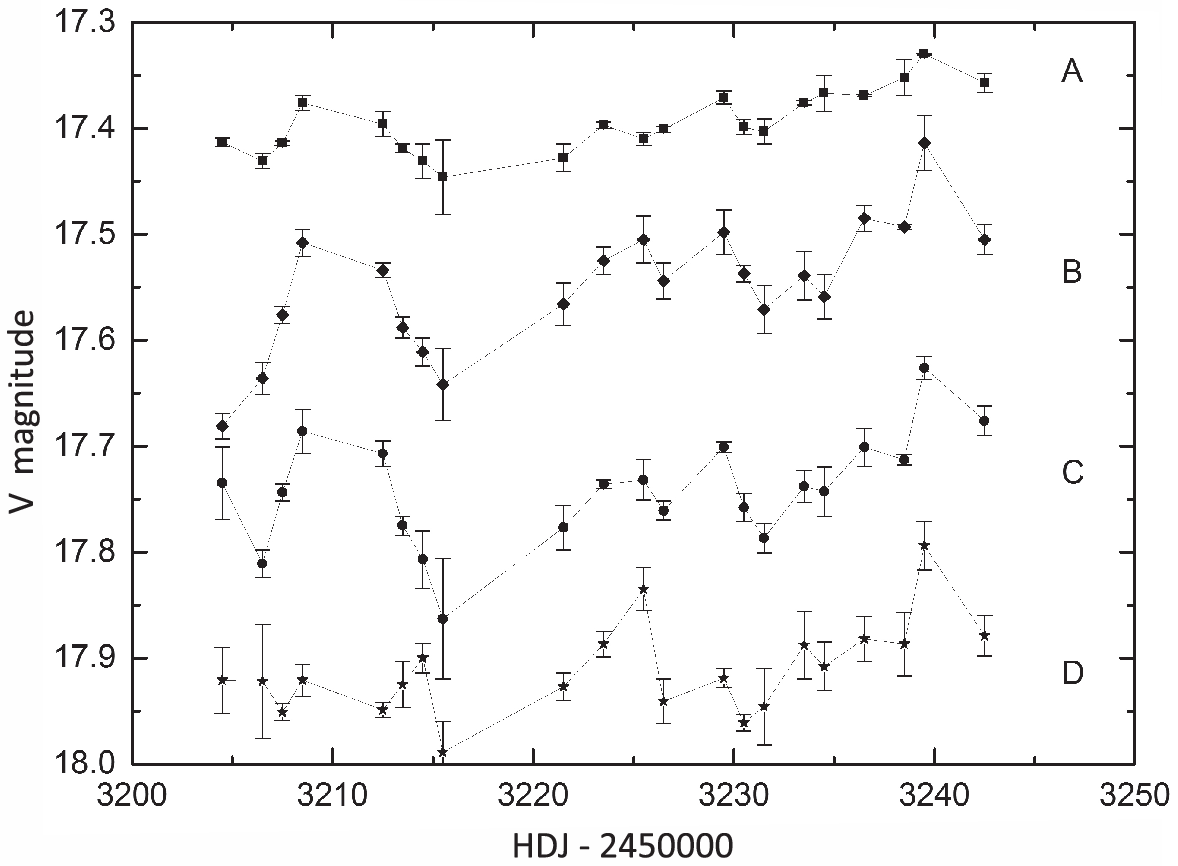,width = 1.03\linewidth}
\caption{The light curves of Q2237+0305 (season 2004) in filter $V$. The magnitudes (along the vertical axis) are shifted arbitrarily for better view; Julian dates - along the horizontal axis.}\label{fig2}
\end{minipage}
\hfill
\begin{minipage}[t]{.47\linewidth}
\centering
\epsfig{file = 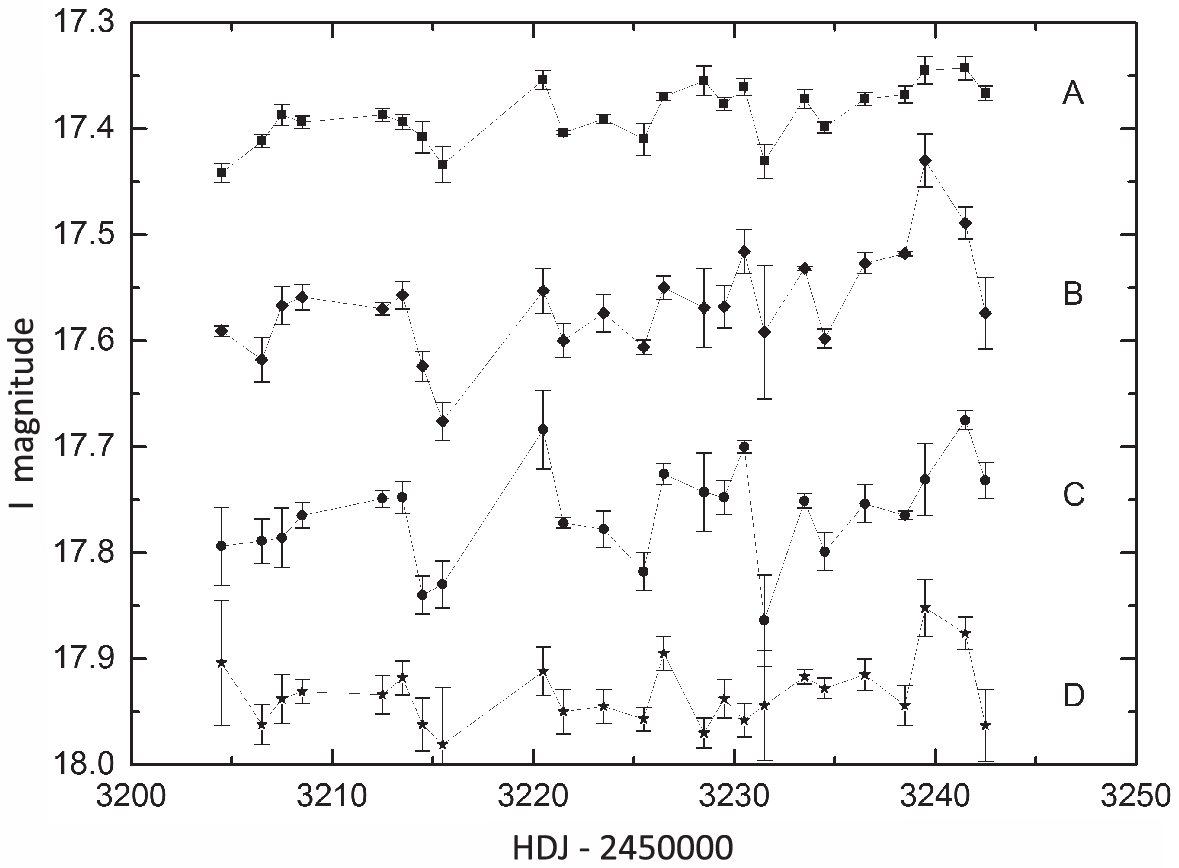,width = 1.03\linewidth}
\caption{The light curves of Q2237+0305 (season 2004) in filter $I$. The magnitudes (along the vertical axis) are shifted arbitrarily for better view; Julian dates - along the horizontal axis.}\label{fig3}
\end{minipage}
\end{figure}

The short-time event under consideration has also been detected in filters $V$ and $I$ (see Fig.~\ref{fig2} and Fig.~\ref{fig3}), excluding component $D$, where the signal-to-noise ratio is too low. As can be seen from a comparison of Figures 1-3, the amplitude of the event is maximal in filter $V$ and decreases towards filters with larger effective wavelengths, $R$ and $I$, with the least amplitude (and the worst detectability) in filter $I$. The dependence of the detected variability amplitudes on wavelength is in agreement with the observations of other authors cited above, (\cite{wilhite08}, \cite{webb00}).

Fig.~\ref{fig1}, Fig.~\ref{fig2} and Fig.~\ref{fig3} also show that the amplitudes of the event are different in different macroimages. In part, this can be caused by the photometry errors, which are not the same in different images. However, this is most probably due to the effects of microlensing we have noted in the Introduction (\cite{barkana97}, \cite{minakov03}).

To evaluate the reliability of the detected short-period variations in the Q2237+0305 quasar intrinsic radiation, we used the statistical F-test. The essence of the test can be found, e.g., in the papers devoted to observations of the rapid variability of quasars (e.g., \cite{bevington03}, \cite{diego10}, \cite{goyal12}, \cite{gopal13}). The F-test is based on the ratio of variance of the process under examination,
$\sigma^2_{obs}$ (an observed light curve in our case) to that of the process we accept
as the reference one, $\sigma^2_{ref}$:
\begin{equation}
F={{\sigma^2_{obs}}\over{\sigma^2_{ref}}}.
\end{equation}

 When the value of $F$ is computed, it should be then compared with the critical value, $F_{crit}$ , which can be found in statistical tables calculated for different significance levels in dependence on the degrees
of freedom $n=N-1$, which are the entries for the tables ($N$
is a number of data points in a given signal record). Application of the F-test to the fragments of light curves shown in Figures 1-3 has demonstrated that the detected flux variations with the amplitudes of up to 0.15-0.2 mag are real with a confidence level of 95 \%.
The amplitudes of the detected event behave in wavelengths just in the same way as is inherent in most quasars subjected to the detailed monitoring, (e.g., \cite{ cristiani97}, \cite{giveon99}, \cite{magdis06},\cite {schmidt12}, \cite {sukanya16}), namely, the amplitudes decrease towards the longer wavelengths.

It is important to note that in spite of several detailed monitoring campaigns of Q2237+0305 (\cite{dudinov11}, \cite{cummings95}, \cite{ostensen96}, \cite{vakulik04}, \cite{udalski06}), no short-period events in the quasar intrinsic flux variations have ever been detected so far. The results of the present work demonstrate the first detection of such events. This allows to expect more accurate time delays determinations, as compared to those ones made earlier by other authors ( \cite{kopt06}, \cite{vakulik06}).

\section*{\sc preliminary results of determining the time delays}
\indent \indent The time delays for the quadruple gravitational lens Q2237 + 0305 were obtained for the first time from the Chandra X-ray observations by Dai \cite{dai03}. Later, the time delays were determined in the optical range \cite{vakulik06} from observations at the Maidanak Observatory in June - October of 2003, and by Koptelova \cite{kopt06} with the use of the OGLE project data for the same time period \cite{udalski06}. It was shown that the time delays in Q2237+0305 are very short. Nevertheless, despite the agreement of these determinations with the theoretic predictions, they have never been used to determine the Hubble constant because of very large errors resulted mainly from the absence of distinct short-period features in the light curves.

To obtain the estimates of the time delays in Q2237+0305 from the light curves shown in Figs. 1-3, we applied the approach reported in our previous work, \cite{tsvetkova16}.  Some unique and useful properties of representing the data of observations by the orthogonal polynomials were applied for realization of the proposed method. For utilizing the properties of the orthogonal-polynomial regressions, we need to construct an orthonormal basis specified at a discrete set of unevenly spaced data points (the dates of observations in our case). This basis can be created with the use of the Gram-Schmidt orthogonalization procedure \cite{korn00}. Any arbitrary set that represents a complete function system specified at the dates of observations can be accepted in this procedure as an initial basis. The system of Legendre polynomials turned out to be the best one in the sense of computational stability in our case.

Representation of the data of observations by the orthogonal polynomials gives the opportunity to eliminate or add any term of the polynomial approximating a light curve, without a necessity to recalculate the rest of the expansion coefficients. This property indicates the way to release the light curves from the component of variability resulted from microlensing effects. It should be noted that eliminating in this way possible contribution from the microlensing effects, we also eliminate the average levels and linear trends in the constituent of the light curves inherent in the intrinsic quasar brightness variations. These low-order terms can be used further to represent differential microlensing without any additional computations.

In choosing the optimal order of the approximating polynomial, we use the following criteria. First, the scatter of the data points relative to the polynomial (approximation error) must tend to approach the error of the original photometric data with an increase of the polynomial order, but must not become smaller. And second, the amplitudes of oscillations near the ends of realizations or in the time periods with sparse sampling must not exceed the magnitude of the photometric errors.

The final step consists in calculating the cross-correlation functions for the pairs of light curves of the components represented by their approximations. To eliminate the edge effects in calculating the cross-correlation functions, a procedure analogous to calculation of a locally normalized discrete correlation function (LNDCF) is used, \cite{lehar92}. Finally, the time shift between the light curves, where the cross-correlation function reaches it maximum is found, and its value is accepted as an estimate of the time delay between a particular pair of macroimages.

\begin{table*}
\centering
\caption{Estimates of the time delays for Q2237+0305 (hours) made from the $R$ light curves in this work.}\label{tab1}
 \vspace*{1ex}
\begin{tabular}{ccccc}
\hline
Time delays, & $R$ (this paper),  & Bar accounted, & SIE lens model, & NSIE+$\gamma$ model,  \\
hours & Dudinov et al. (2011)  & Schmidt et al. (1998) &  Wertz et al. (2014)  &  Wertz et al. (2014) \\
\hline
\hline
$\Delta t_{AB}$ &  4.5$\pm$1.9     &   2.0    & 2.6$\pm$0.7     & 2.51$\pm$0.52 \\
$\Delta t_{AC}$ & -14.22$\pm$1.66  &   -16.2  & -17.97$\pm$0.7  & -18.01$\pm$0.62  \\
$\Delta t_{AD}$ & -2.73$\pm$4.61   &   -4.9   & -5.38$\pm$0.7   & -5.41$\pm$0.43 \\
$\Delta t_{BC}$ & -19.00$\pm$1.56  & (-18.27) &    (-20.57)     & (-20.52)    \\
$\Delta t_{BD}$ & -2.97$\pm$3.96   &  (-6.9)  &    (-7.98)      &    (-7.92) \\
$\Delta t_{CD}$ &  19.31$\pm$2.76  &  (11.3)  &  12.59$\pm$0.7 &  12.6$\pm$0.58 \\
\hline
\end{tabular}
\end{table*}

The values of the time delays for image pairs $AB$, $AC$, $AD$, $BC$, $BD$ and $CD$ of Q2237+0305 obtained from the $R$ light curves by Dudinov et al.  \cite{dudinov11}, are presented in Table 1, (see the first column) together with the most recent model predictions for this quadruply lensed quasar. The last three columns correspond to model predictions taken from the most recent works, with the values for $\Delta t_{BC}$, $\Delta t_{BD}$, $\Delta t_{CD}$ (in brackets) calculated from the original $\Delta t_{AB}$, $\Delta t_{AC}$, $\Delta t_{AD}$ with the use of the known relationship between the time delays in quadruple lenses $\Delta t_{ij}-\Delta t_{ik}=\Delta t_{kj}$.  The measured time delays are seen to be consistent with those calculated in \cite{wertz14} with the use of the most adequate models for the gravitation potential of the lensing galaxy and most accurate astrometry. The least relative errors of the time delays are achieved for pairs $AC$, $BC$ and $CD$ (12 \%, 8 \% and 14 \%, respectively).
Processing of the $V$ and $I$ light curves gives rather diverging results, which need further analysis. For the $V$ light curves one of the explanations may be connected with the presence of the broad emission line within the filter passband.

\section*{\sc results and conclusions}
\indent \indent In spite of decades of the comprehensive studies, variability of quasars still remains to be rather poorly understood phenomenon. Detection of the rapid variability, at the time scale of several days, made for Q2237+0305 for the first time since its discovery in 1985  \cite{huchra85}, is important  in two aspects. Firstly, it may help to distinguish between various mechanisms of the quasar variability, and secondly, the light curves with the short-period features are anticipated to provide more accurate time delays as compared to those obtained earlier. As is seen from Table 1 of the present work, our estimates of the time delays are not only well consistent with the model predictions, but also demonstrate, of all the existing determinations, the lowest uncertainties, of the order of a few hours.

\section*{\sc acknowledgement}
\indent \indent We would like to thank O.I. Bugaenko for the unceasing interest to the work, efficient and encouraging discussions, which have hopefully served better quality of the paper.


\end{document}